\documentclass{IOS-Book-Article}
\usepackage[sort&compress,numbers]{natbib}

\usepackage{hyperref}

\usepackage{xcolor}
\usepackage{graphicx}
\usepackage[shortlabels]{enumitem}

\usepackage{mathptmx}

\def\hb{\hbox to 10.7 cm{}}
\newcommand{\msun}{{\rm M}_{\odot}}
\begin{document}

\pagestyle{headings}
\def\thepage{}

\begin{frontmatter}              

\title{Inference-optimized AI and \\ high performance computing for \\ gravitational wave detection at scale}


\author[A,B,D]{\fnms{Pranshu} \snm{Chaturvedi}%
\thanks{Corresponding Author; email:
pranshu3@illinois.edu \& pchaturvedi@anl.gov}},
\author[A,C,D]{\fnms{Asad} \snm{Khan}},
\author[C,D]{\fnms{Minyang} \snm{Tian}},
\author[A,C,E]{\fnms{E.~A.} \snm{Huerta}}
and
\author[B]{\fnms{Huihuo} \snm{Zheng}}


\address[A]{Data Science and Learning Division, Argonne National Laboratory, Lemont, Illinois 60439,
USA}
\address[B]{Department of Computer Science, University of Illinois at Urbana-Champaign, Urbana, Illinois 61801, USA}
\address[C]{Department of Physics, University of Illinois at Urbana-Champaign, Urbana, Illinois 61801, USA}
\address[D]{National Center for Supercomputing Applications, University of Illinois at Urbana-Champaign, Urbana, Illinois 61801, USA}
\address[E]{Department of Computer Science, University of Chicago, Chicago, Illinois 60637, USA}
\address[F]{Leadership Computing Facility, Argonne National Laboratory, Lemont, Illinois 60439,
USA}

\begin{abstract}
We introduce an ensemble of artificial intelligence 
models for gravitational wave detection that we 
trained in the Summit supercomputer using 
32 nodes, equivalent to 192 NVIDIA V100 GPUs, within 
2 hours. Once fully trained, we optimized these models 
for accelerated inference using \texttt{NVIDIA TensorRT}. 
We deployed our 
inference-optimized AI ensemble in the 
ThetaGPU supercomputer at Argonne Leadership 
Computer Facility to conduct distributed 
inference. Using the entire ThetaGPU supercomputer, 
consisting of 20 nodes each of which 
has 8 NVIDIA A100 Tensor Core GPUs and 2 AMD Rome CPUs,
our \texttt{NVIDIA TensorRT}-optimized AI ensemble 
processed an entire month of advanced LIGO data (including 
Hanford and Livingston data streams)
within 50 seconds. Our 
inference-optimized AI ensemble retains the 
same sensitivity of traditional AI models, namely, 
it identifies 
all known binary black hole 
mergers previously identified in this advanced LIGO 
dataset and 
reports no misclassifications, 
while also providing a \(3X\) inference speedup 
compared to traditional artificial intelligence models. 
We used time slides to quantify the 
performance of our AI ensemble to process up to 5 
years worth of advanced LIGO data. In this 
synthetically enhanced dataset, our AI ensemble reports an average of 
one misclassification for every month of searched advanced LIGO data. 
We also present the 
receiver operating characteristic
curve of our AI ensemble using this 5 year long advanced LIGO dataset.
This approach provides the required tools to conduct 
accelerated, AI-driven gravitational wave detection 
at scale.
\end{abstract}

\end{frontmatter}


\section{Introduction}

The international network of ground-based gravitational 
wave interferometers---advanced 
LIGO~\cite{bbhswithligo:2016, LIGOScientific:2016emj}, advanced Virgo~\cite{Virgo:2015,aVirgo:2020} 
and Kagra~\cite{KAGRA:2020}---have completed three observing runs, reporting 
the detection of tens of gravitational wave sources~\cite{LIGOScientific:2021djp}. 
Within the next decade, these scientific facilities will 
usher in the era of precision gravitational wave astrophysics, 
shedding new light into the astrophysical properties of 
gravitational wave sources, likely formation scenarios, 
and the nature of the environments where 
they reside~\cite{LIGOScientific:2021psn}. 
We have already witnessed the 
transformational power of gravitational wave astrophysics 
in fundamental physics, cosmology, chemistry and 
nuclear physics~\cite{LIGOScientific:2021aug,kiloGW170817:2017,2019Natur.568..469M,bnsdet:2017,grb:2017ApJ,Tan:2020ics,Yunes:2016jcc,LIGOScientific:2020tif,Mooley_nat}. These are 
only a few glimpses of the 
scientific revolution that may take place within the next decade~\cite{Reitze:2021gzo,Couvares:2021ajn,Punturo:2021ryo,Kalogera:2021bya,McClelland:2021wqy} 
if we translate the data deluge to be 
delivered by gravitational wave detectors into the required 
elements to enable scientific discovery at scale.

Realizing the urgent need to develop novel 
frameworks for scientific discovery that adequately address 
challenges brought about by the big data revolution, 
and acknowledging that many disciplines are undergoing 
similar transformations thereby increasing the demand 
on already oversubscribed computational resources, 
scientists across the 
world are eagerly developing the next generation of 
computing frameworks and signal processing tools that will 
enable the realization of this research program~\cite{Nat_Rev_2019_Huerta}.

Over the last few years, it has become apparent that 
the convergence of 
artificial intelligence (AI) and innovative computing provides 
the means to tackle computational grand challenges that 
have been exacerbated with the advent of 
large scale scientific facilities, and which will not 
be met by the 
ongoing deployment of exascale HPC systems alone~\cite{HPCUSE,2020arXiv200308394H}. 
As described in recent reviews~\cite{cuoco_review,huerta_book}, 
AI and high performance 
computing (HPC) as well as edge computing have been 
showcased to enable gravitational wave detection with the 
same sensitivity than template-matching algorithms, but 
orders of magnitude faster and at a fraction of the 
computational cost. At a glance, 
recent AI applications 
for gravitational wave astrophysics includes 
classification or signal detection~\cite{geodf:2017a,George:2018PhLB,2018GN,2020arXiv200914611S,Lin:2020aps,Wang:2019zaj,Fan:2018vgw,Li:2017chi,Deighan:2020gtp,Miller:2019jtp,Krastev:2019koe,2020PhRvD.102f3015S,Dreissigacker:2020xfr,Adam:2018prd,Dreissigacker:2019edy,2020PhRvD.101f4009B,2021arXiv210810715S,2021arXiv210603741S,2021arXiv210812430G}, signal denoising and data cleaning~\cite{shen2019denoising,Wei:2019zlc,PhysRevResearch.2.033066,Yu:2021swq}, regression or parameter estimation~\cite{Shen:2019vep,Gabbard:2019rde,Chua:2019wwt,Green:2020hst,Green:2020dnx,2021arXiv210612594D,2021arXiv211113139D,Khan_HOM}, 
accelerated waveform production~\cite{Khan:2020fso,PhysRevLett.122.211101}, signal forecasting~\cite{2021PhRvD.103l3023L,khan_huerta_zheng_forecast}, 
and early warning systems for 
gravitational wave sources that include matter, such as binary neutron stars or black hole-neutron star systems~\cite{Wei_quantized,Wei:2020sfz,2021PhRvD.104f2004Y}.

In this article, we build upon our recent work 
developing AI frameworks for production scale 
gravitational wave detection~\cite{2021PhLB..81236029W,huerta_nat_ast}, 
and introduce an 
approach that consists of optimizing AI models 
for accelerated inference, 
leveraging \texttt{NVIDIA TensorRT}~\cite{trrt}. We describe how we 
deployed our \texttt{TensorRT} AI ensemble in the 
ThetaGPU supercomputer at Argonne Leadership Computing 
Facility, and developed the required software to 
optimally distribute inference using up to 20 nodes, 
which are equivalent to 160 NVIDIA A100 Tensor 
Core GPUs. We quantified 
the sensitivity and computational efficiency of 
this approach by processing the entire month of 
August 2017 of advanced LIGO data (using both 
Hanford and Livingstone datasets). Our analysis 
indicates that with 
our proposed approach, we are able to process these datasets 
within 50 seconds using 20 nodes in the ThetaGPU 
supercomputer at Argonne Leadership Computing Facility. Most 
importantly, we find that these optimized models retain the 
same sensitivity of traditional AI models, since they are 
able to identify all binary black hole mergers in this 
month-long dataset, while also reporting no 
misclassifications, and reducing time-to-insight 
by up to \(3X\) compared to traditional AI models~\cite{huerta_nat_ast}.

This article is organized as follows. Section~\ref{sec:met} 
describes the approach we followed to train our AI 
models, optimize them for accelerated inference, and then 
combined them to search for gravitational waves as 
an ensemble. We also describe the advanced LIGO 
datasets used for training, validation and testing. We 
summarize our findings in Section~\ref{sec:res}. 
We outline future directions of work in Section~\ref{sec:end}.

\section{Materials and Methods}
\label{sec:met}

Here we describe the AI architecture used for 
these studies, the modeled waveforms and 
advanced LIGO data used to train and test 
a suite of AI models. We then describe the 
procedure to optimize an ensemble of AI 
models for accelerated AI inference, and the approach 
followed to deploy this AI ensemble 
in the ThetaGPU supercomputer to optimally 
search for gravitational waves in advanced 
LIGO data at scale.

\noindent \textbf{Modeled waveforms} In this study we 
consider binary black hole mergers, and produce 
synthetic signals that describe them with the 
\texttt{SEOBNRv3} waveform model~\cite{seobnrv3} that is 
available in the open source \texttt{PyCBC} 
library~\cite{pycbc_library}. We densely sample 
a parameter space that comprises black hole binaries with 
mass-ratios \(1\leq q \leq 5\), individual 
spins \(s^z_{\{1,2\}}\in[-0.8,\,0.8]\), and 
total mass \(M\in[5\msun,\,100\msun]\). We 
used a training dataset of over 1,136,415 waveforms, and a 
validation and testing datasets of over 230k waveforms,  
sampled at 4096 Hz, to create a
suite of AI models in the Summit supercomputer. 

\noindent \textbf{Advanced LIGO data} We used 
advanced LIGO data available through the 
Gravitational Wave Open Science Center~\cite{Vallisneri:2014vxa}. 
The three data 
segments we consider have initial GPS times $1186725888$, 
$1187151872$, and $1187569664$, and are 4096 seconds long. 
Each of these segments include both Hanford and 
Livingstone data, and do not include known 
gravitational wave signals. 
 
\noindent \textbf{Data preparation} We used 
advanced LIGO data to compute power spectral density (PSDs) 
estimates using open source software available at the 
Gravitational Wave Open Science Center. 
We used these PSDs to whiten both modeled waveforms 
and advanced LIGO strain data, which are then 
linearly combined to simulate a wide range of 
astrophysical scenarios, covering a broad range of 
signal-to-noise ratios. Following best practices 
for the training of AI models, we normalized the standard 
deviation of training data that contain both signals and 
noise to one. We combined our set of 
1,136,415 modeled waveforms with advanced LIGO noise 
by randomly sampling 1 second long contiguous data samples. 
To be precise, since we use advanced LIGO data sampled at 4096 Hz, 
this means that a 1 second long segment may be described as a set 
of continuous samples covering the range 
\([i_1, \dots, i_{4096}]\). In the same vein, another noise 
realization may be given by the samples 
\([i_{520}, \dots, i_{4596}]\), etc. This means that in any of 
the 4096 second long advanced LIGO data segment we use for 
training, we could draw \(4096\times4096-4096+1\) contiguous, 
1 second long noise segments. Since we consider \(3\times4096\,\textrm{second}\) 
long advanced LIGO data 
segments per detector, then it follows that we have at our disposal 
about 50M noise realizations per detector. Notice, however, that 
each input that we feed into the net is distinct to each other. 
This is because each whitened waveform has unique astrophysical 
parameters, \((M, q, s_1^z, s_2^z)\), and is linearly combined 
with a whitened noise realization that simulates a variety of 
signal to noise ratio scenarios. On the other hand, 
we actually find that the number 
of noise realizations we use for training per detector 
is given by \(\#\, \textrm{of training iterations} 
\times \mathrm{batch size} \). In our case \(\#\, \textrm{of training iterations} \rightarrow2,556,933\) and \(\mathrm{batch size} \rightarrow 16\). In other words, we use about 40M noise realizations to 
produce AI models that exhibit strong convergence and 
optimal performance for gravitational wave detection.

\begin{figure}[h!]
\includegraphics[width=0.9\linewidth]{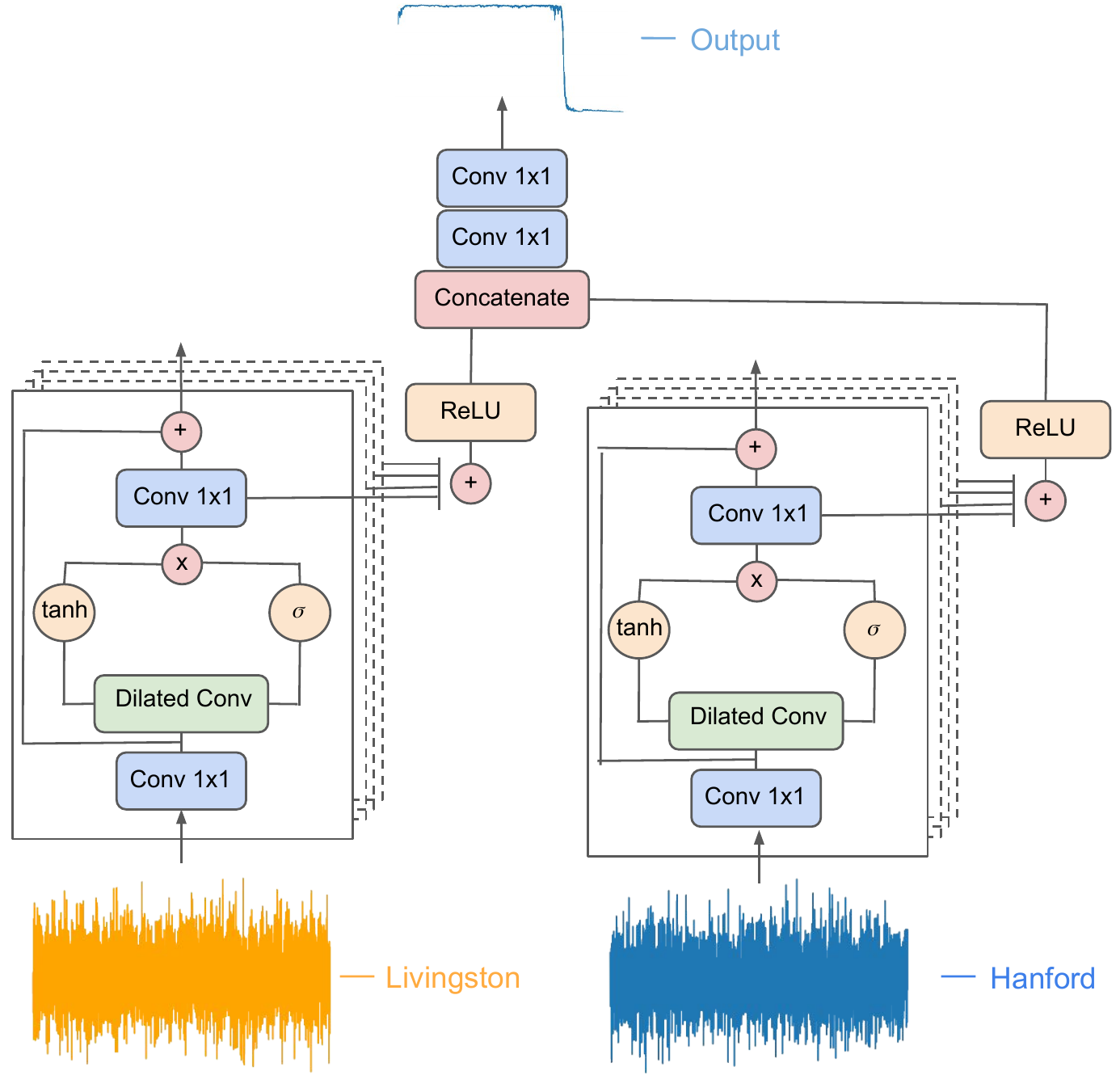}
\caption{\textbf{AI Architecture}. Modified \texttt{WaveNet} 
model used for gravitational wave detection. Each branch 
processes concurrently one of the two advanced LIGO data 
streams---Hanford or Livingston. The output of the 
two branches is then concatenated 
and fed into a pair of convolutional layers whose output indicates 
at each time step whether the input advanced LIGO data 
contains ``noise'' or a ``waveform''. 
}
\label{fig:ai_model}
\end{figure}

\noindent \textbf{AI architecture} We designed a modified \texttt{WaveNet}~\cite{2016wavenet} 
architecture that takes in advanced LIGO strain, both from 
Livingston and Hanford, sampled at 4096Hz. The two outputs 
of these 
models (one for each advanced LIGO strain data) are combined 
and then fed into a set two convolutional layers whose output 
consists of a classification probability for each time step. 
The AI architecture used in these studies is depicted in 
Figure~\ref{fig:ai_model}.

\begin{figure}[h!]
\includegraphics[width=\linewidth]{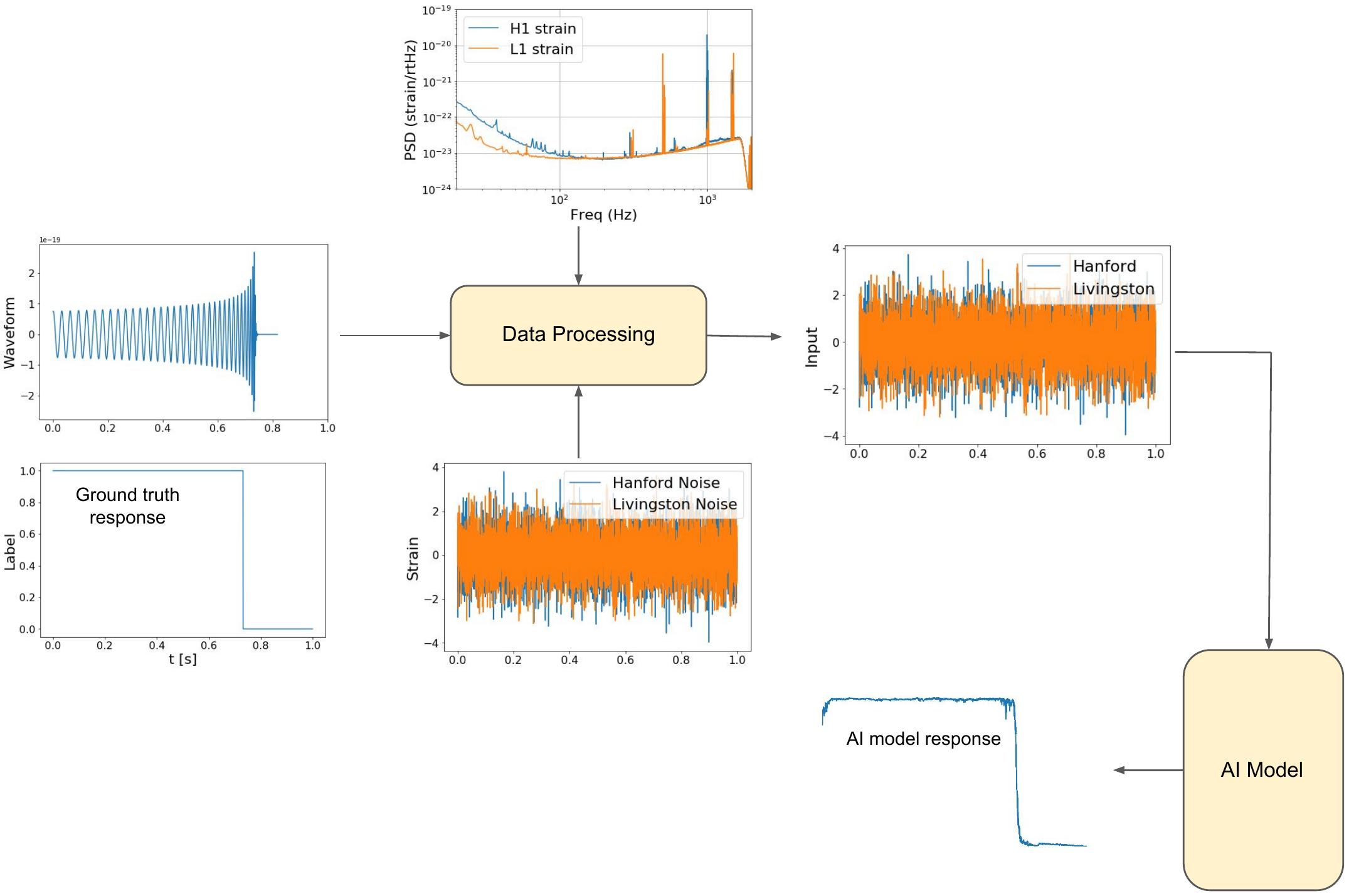}
\caption{\textbf{Model creation}. Methodology used for data 
curation, model training and testing.
}
\label{fig:curation}
\end{figure}

\noindent \textbf{AI ensemble construction} During training, 
the ground-truth labels are curated such that 
each time step after the merger of a given modeled waveform 
is classified as ``noise'', whereas all the preceding 
time steps are classified as ``waveform''. We used 
the AI architecture described above and trained a suite of tens 
of AI models with the Summit supercomputer. We used the same 
architecture but allowed for random initialization of weights. 
Each model 
was trained using 32 Summit nodes, equivalent to 
192 NVIDIA V100 GPUs.
We then picked a sample of the best ten models 
and quantified their 
classification accuracy. We did so by leveraging the feature 
we encoded in the models to flag the transition between 
``noise'' and ``waveform'', which corresponds to the location 
of the merger of a binary black hole merger. Thereafter, 
we took the output of these models and post-processed it with 
the \texttt{find\_peaks} algorithm, a \texttt{SciPy}'s 
off-the-shelve tool, to accurately identify the location of 
these mergers. Finally, we created several combinations of 
these models and quantified the optimal ensemble that 
maximized classification accuracy while also reducing 
the number of false positives in minutes-, 
hours-, weeks- and a month-long advanced LIGO strain 
datasets. This entire methodology, from data curation to model 
training and testing is schematically presented in 
Figure~\ref{fig:curation}. Having 
identified an optimal AI ensemble, we 
optimized it for accelerated inference using 
\texttt{TensorRT}.

\noindent \textbf{Optimization with \texttt{NVIDIA TensorRT}.} 
To further reduce time-to-insight with our AI ensemble, 
we converted our existing AI models, which were 
originally created in \texttt{TensorFlow 1} to 
\texttt{TensorRT 8} engines. The first step in the conversion process requires us to convert our \texttt{HDF5} 
files containing the architecture and weights into 
the \texttt{TensorFlow SavedModel} format. We then 
make use of \texttt{tf2onnx}~\cite{onnx}, an 
open-source tool for converting \texttt{SavedModels} to 
the \texttt{Open Neural Network Exchange (ONNX)} format~\cite{onnxlib}. Next, we created a script 
to describe and build our \texttt{TensorRT} engines 
and accordingly specified the following parameters: 
the maximum amount of memory that can be allocated by the 
engine, which was set to 32 GB (NVIDIA A100 GPUs have 40GB 
of memory), allowed half-precision (FP16) 
computation where possible, 
the input dimensions of the model including the batch size 
\((1024, 4096, 2)\), the output of the model 
\((1024, 4096, 1)\), and a flag that allows the built engine 
to be saved so that the engine will not have to be 
reinitialized in subsequent runs. \texttt{TensorRT} 
applies a series of optimizations to the model by running 
a GPU profiler to find the best GPU kernels to use 
for various neural network computations, applying 
graph optimization techniques to reduce the number 
of nodes and edges in a model such as layer fusion, 
quantization where appropriate, and more. We found 
that the \texttt{TensorRT} ensembles allowed us to 
increase the batch size from 256 to 1024 due to the 
compressed architecture generated by \texttt{TensorRT} 
and found an overall average speedup of \(3X\) 
when using the entire ThetaGPU systems for accelerated 
gravitational wave inference.

\noindent \textbf{Inference-optimized AI ensemble 
deployment in ThetaGPU} 
We developed software to optimally process 
advanced LIGO data using the ThetaGPU 
supercomputer. We quantified 
the performance of this approach using 1, 2, 4, 8, 12, 16 
and 20 nodes to demonstrate strong scaling. 
Parallelization was done with \texttt{mpi4py} built 
on \texttt{OpenMPI 4}. Each GPU, in every ThetaGPU node, 
acts as one \texttt{MPI} process in our parallel 
inference script. 

\section{Results}
\label{sec:res}

We present three main results: statistical 
analysis, noise anomaly processing, and computational efficiency 
of our AI-driven search.

\subsection{Event Detection}

\begin{figure}[h!]
\centerline{
\includegraphics[width=.9\linewidth]{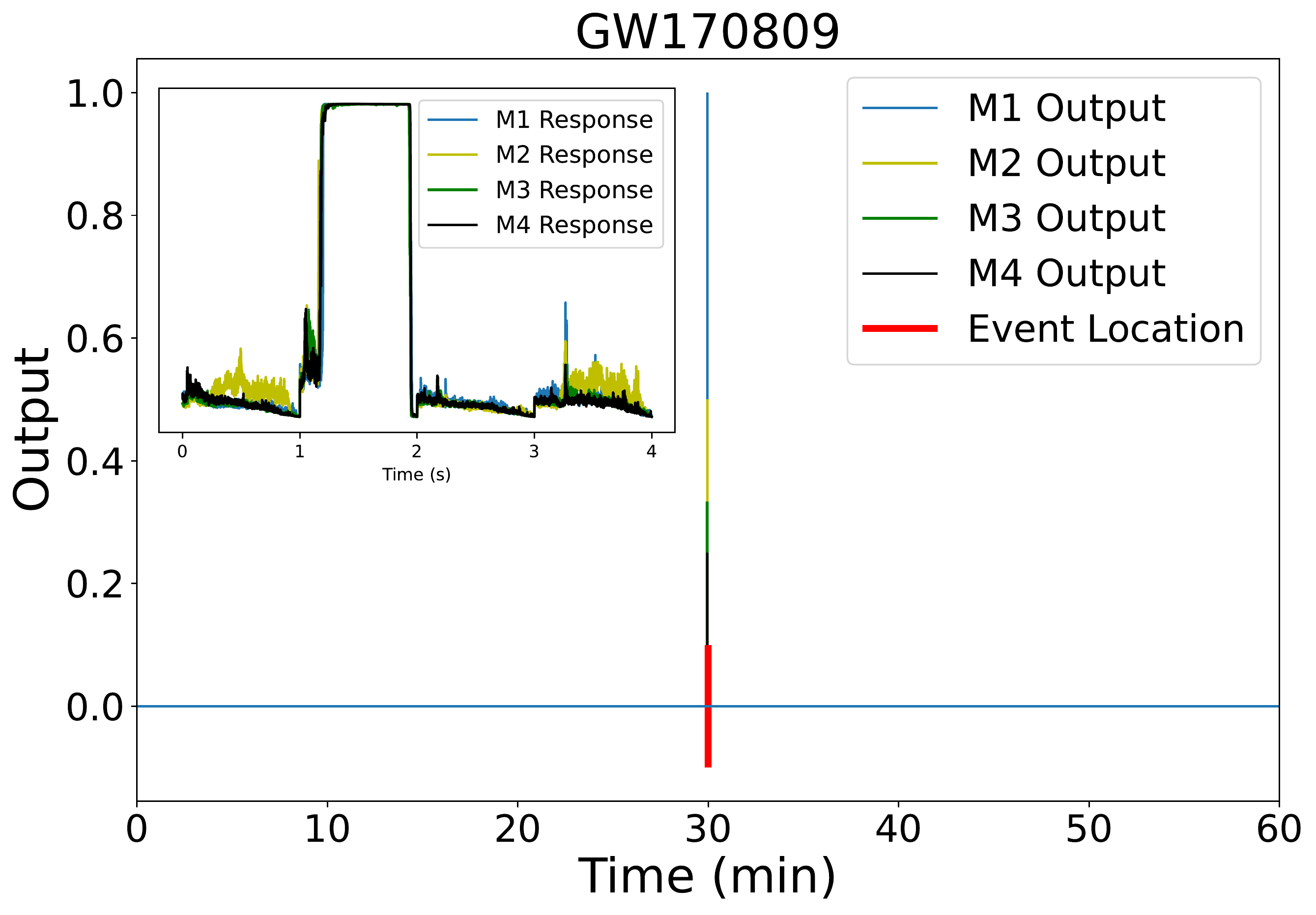}
}
\centerline{
\includegraphics[width=0.9\linewidth]{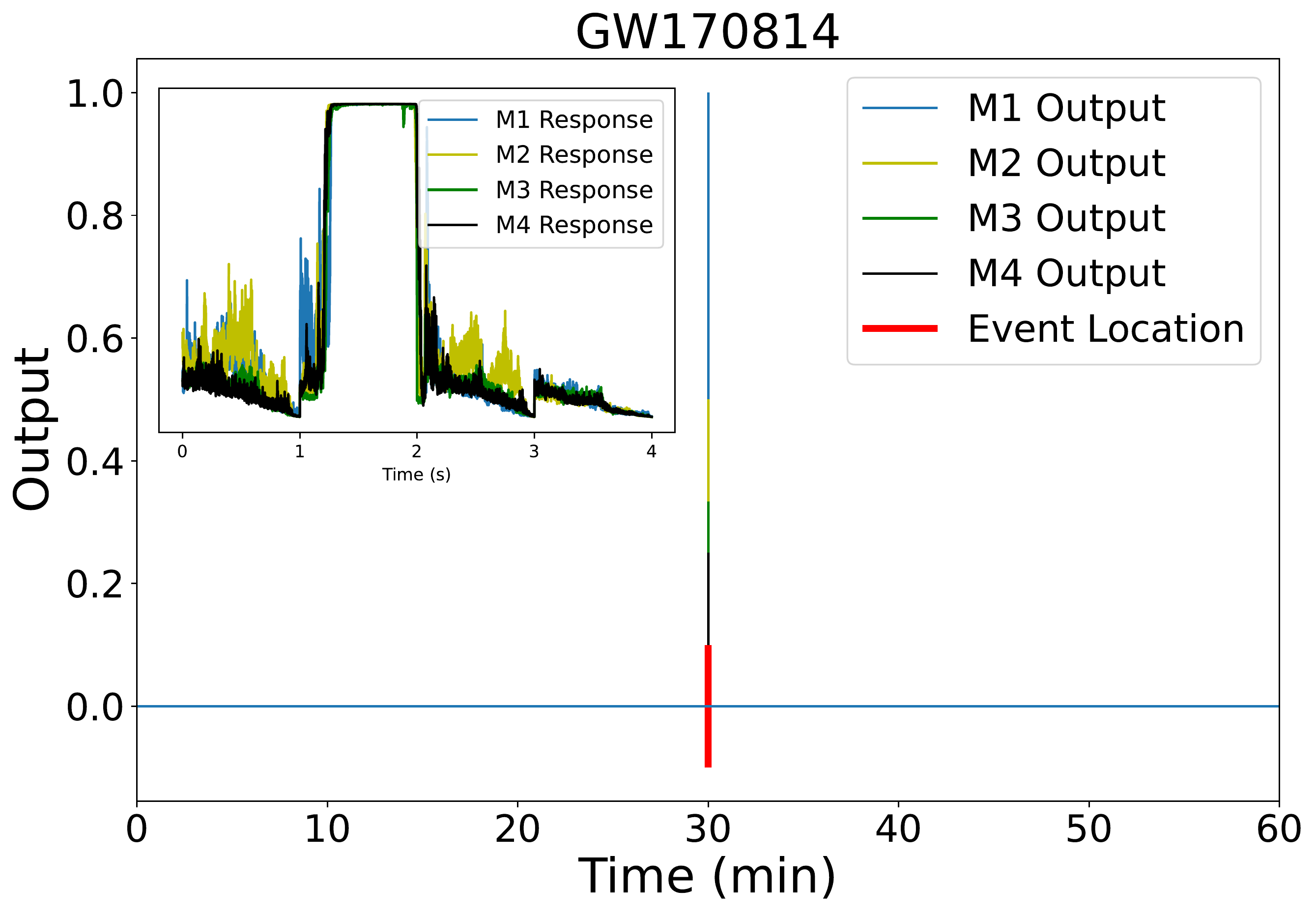}
}
\caption{\textbf{Event Detection} Output of the 4 individual AI models in our ensemble upon 
processing 1 hour long advanced LIGO data that contains the events GW170809 (top panel) and GW170814 
(bottom panel). The insets in both panels show the distinct, sharp response that is common among all AI models 
when they identify a real signal.
}
\label{fig:all_events_I}
\end{figure}

\begin{figure}[h!]
\centerline{
\includegraphics[width=0.9\linewidth]{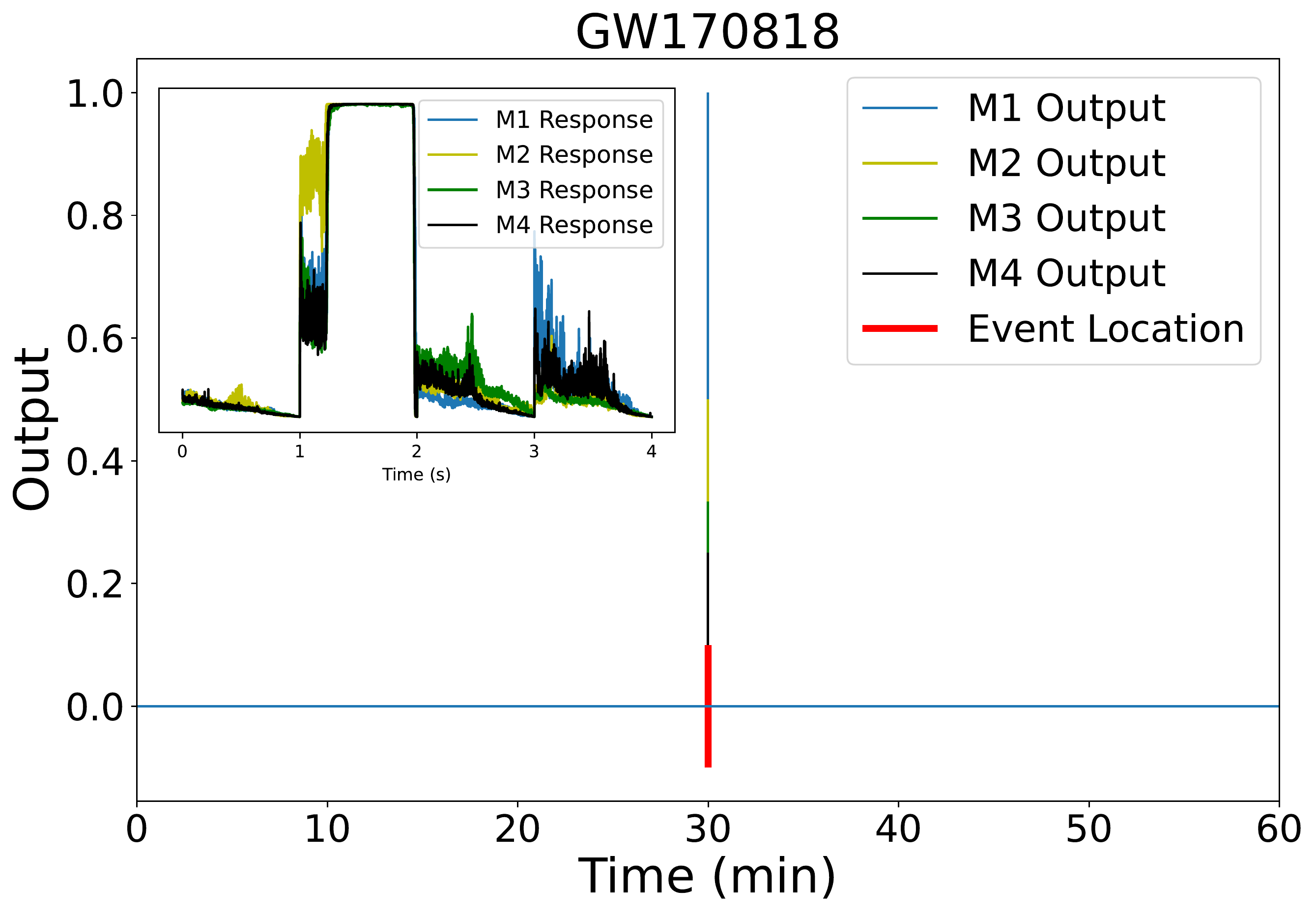}
}
\centerline{
\includegraphics[width=0.9\linewidth]{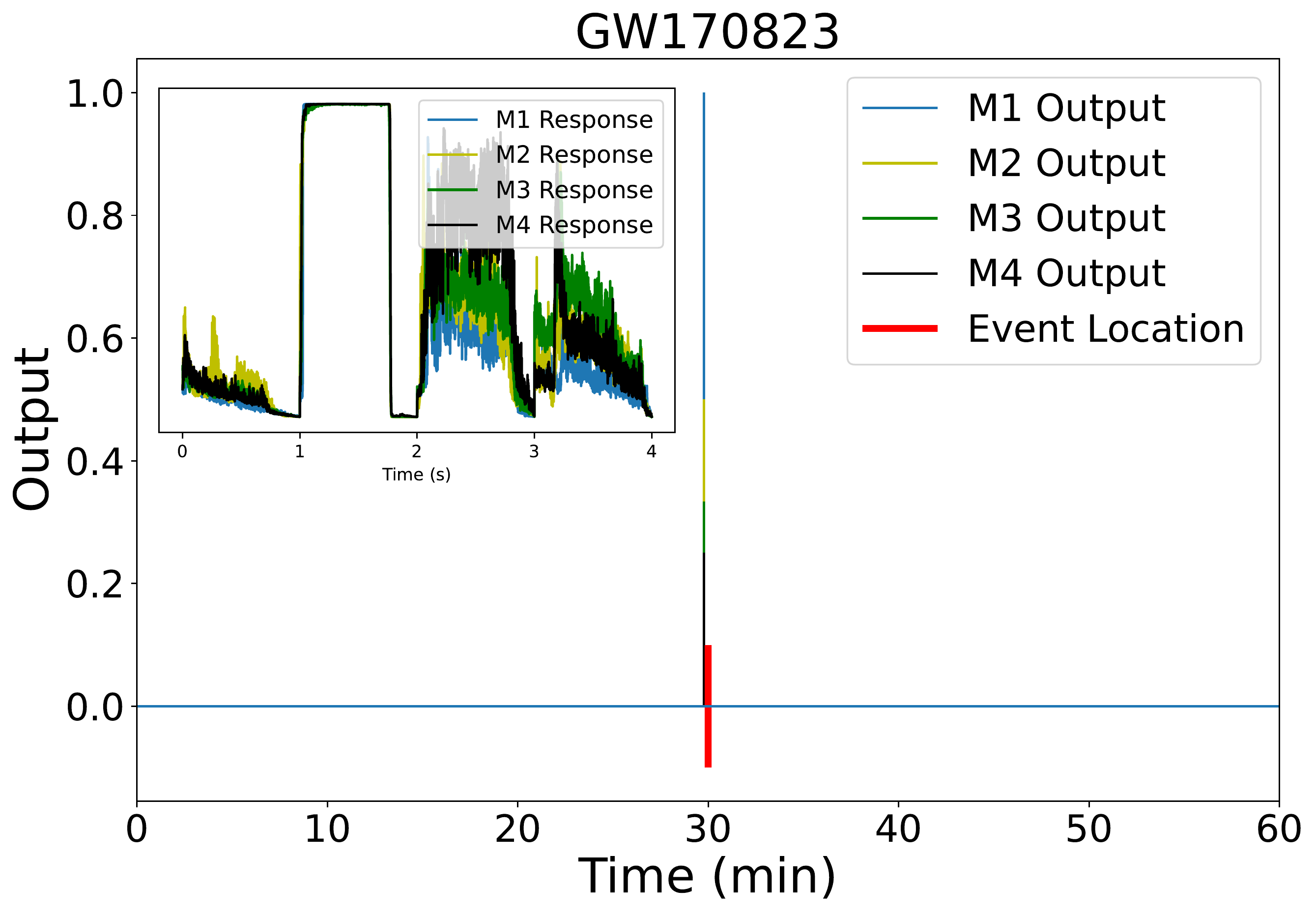}
}
\caption{\textbf{Event Detection} As Figure~\ref{fig:all_events_I}, but now for GW170818 
(top panel) and GW170823 (bottom panel).
}
\label{fig:all_events_II}
\end{figure}

We used our 
inference-optimized AI ensemble to process hours-, 
days-, weeks-, and a month-long advanced LIGO dataset. 
We found that this AI ensemble was able to identify 
all binary black hole mergers reported throughout the 
second observing run that covered the month of
August 2017. Figures~\ref{fig:all_events_I} 
and~\ref{fig:all_events_II} show the distinct,
sharp response of each of our AI models in the ensemble when they identify 
real gravitational wave signals. Notice also that the individual models 
report no other noise trigger of importance within one hour of 
data of these four events GW170809, 
GW170814, GW170818, and GW170823. 
While Figures~\ref{fig:all_events_I} 
and~\ref{fig:all_events_II}  show the response of our 
AI ensemble in the vicinity of these events, we conducted 
a systematic analysis for all the noise triggers reported 
by the ensemble upon processing the entire month of 
August 2017. Triggers that were reported by all AI 
models in the ensemble, and which were coincident within 
a time window of 0.5 seconds were flagged as 
gravitational wave events. Our analysis only reported 
four noise triggers of that nature, namely, GW170809, 
GW170814, GW170818, and GW170823.

\subsection{Noise Anomaly Processing}

\begin{figure}[h!]
\centerline{
\includegraphics[width=0.85\linewidth]{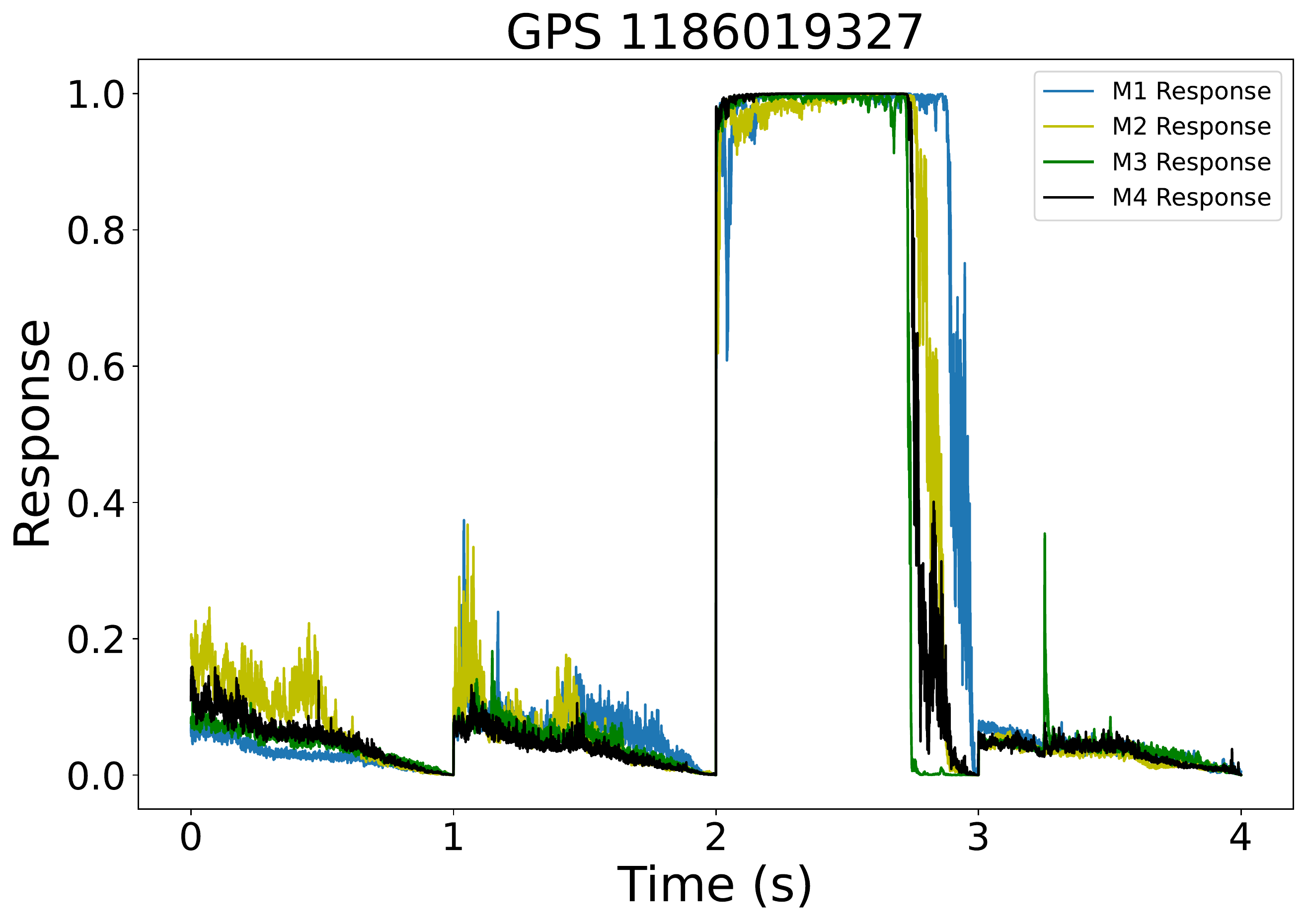}
}
\centerline{
\includegraphics[width=0.85\linewidth]{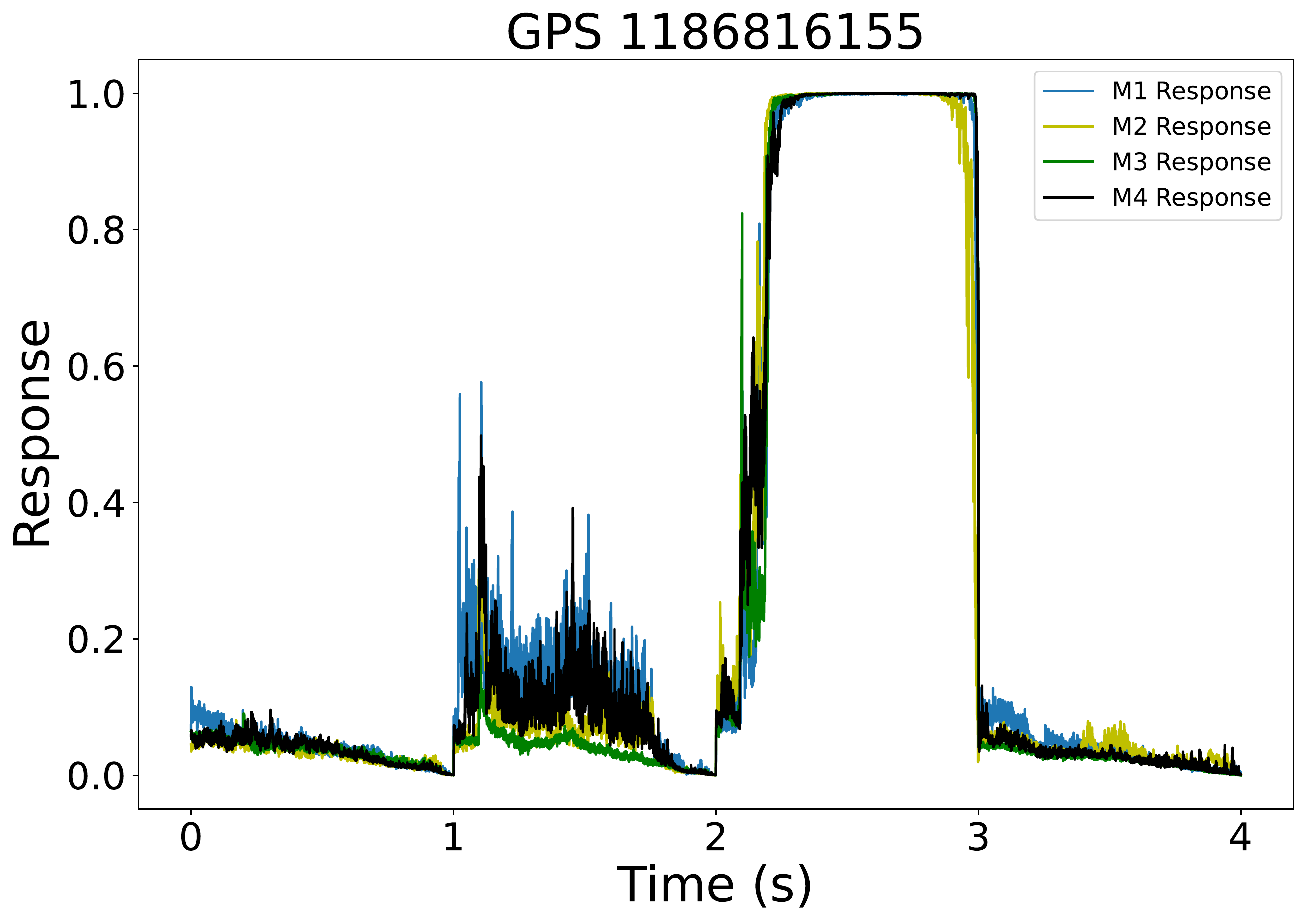}
}
\caption{textbf{Noise anomalies}  Response of our AI ensemble to real glitches 
located at GPS times 1186019327 (top panel) and 1186816155 (bottom panel).
}
\label{fig:glitches}
\end{figure}

\noindent We quantified the performance 
of our AI ensemble to 
discard noise anomalies. To do so, we considered three 
real glitches in August 2017, namely those 
with GPS times 1186019327 and 1186816155. 
In Figure~\ref{fig:glitches} we 
show the response of our AI ensemble to each of these 
noise triggers. We notice that the individual AI models 
in the ensemble do not agree on the nature of these 
noise triggers, and thus we readily discard them as 
events of interest. Key features that our \texttt{find\_peaks} algorithm 
utilizes to discard these events encompass the jaggedness and inconsistent 
widths of these peaks. Since our AI ensemble only 
identified actual gravitational wave events as 
relevant noise triggers throughout August 2017, we 
conclude that our AI ensemble was capable of discarding all
other glitches in this one month long data batch.
    
\subsection{Statistical Analysis}

\noindent We have quantified the performance 
of our AI classifiers by going beyond the one month worth of data 
that we used in the previous section for event detection and noise anomaly 
rejection. To do this, we use time slides to synthetically enhance the 
month long August 2017 advanced LIGO dataset. Using the approach 
described in~\cite{2021arXiv210810715S}, we produced datasets that span 
between 1 to 5 years of advanced LIGO data. Our findings show that our 
AI ensemble reports, on average, about 1.3 false positives per month. 
Specifically, we found that the number of false positives for each 
time-shifted dataset are:

\begin{itemize}[nosep]
\item 1 year worth of data. 22 false positives
\item 2 years worth of data. 35 false positives 
\item 3 years worth of data. 53 false positives 
\item 4 years worth of data. 68 false positives
\item 5 years worth of data. 79 false positives
\end{itemize}

\begin{figure}[h!]
\centerline{
\includegraphics[width=\linewidth]{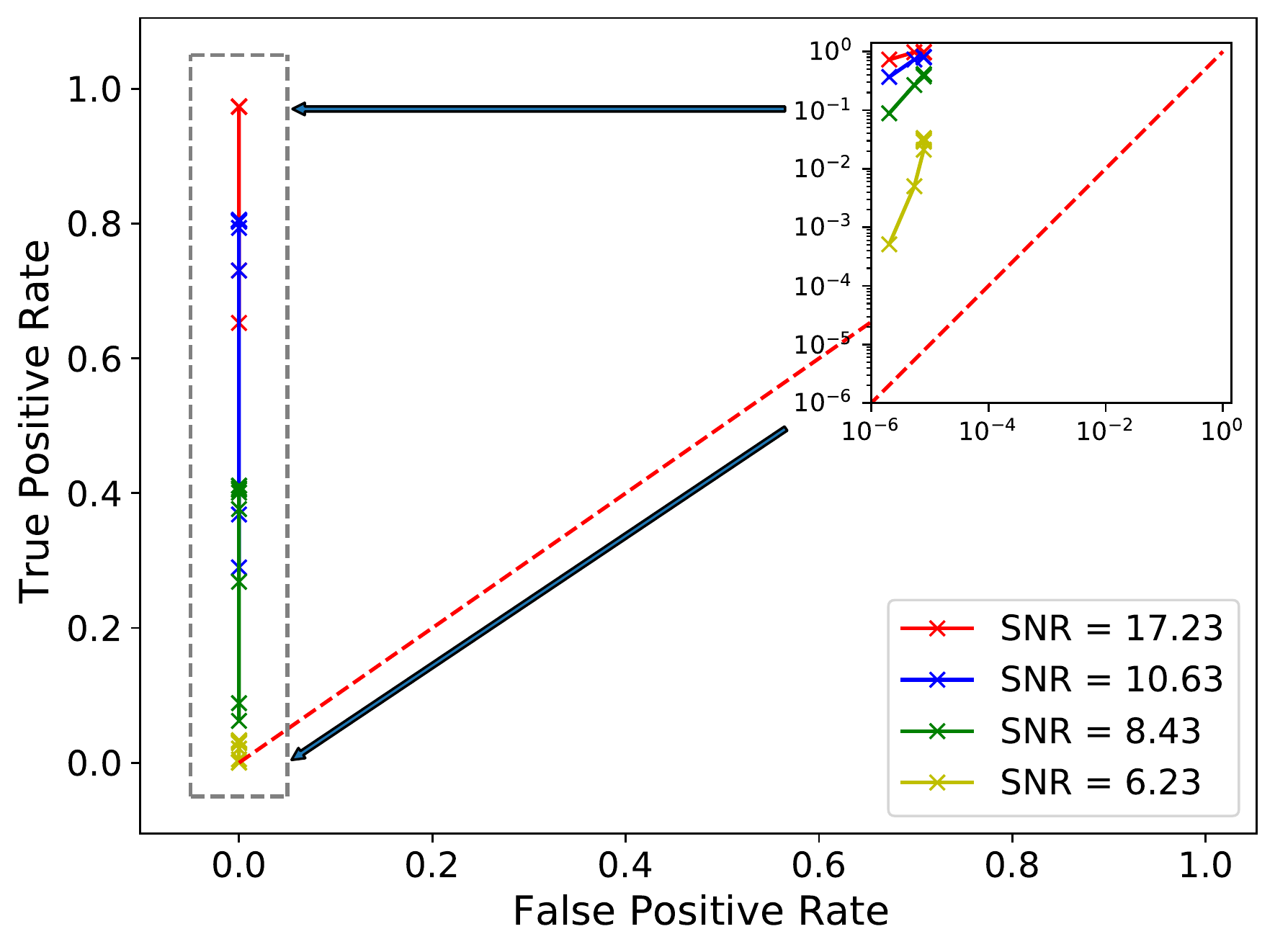}
}
\caption{\textbf{Receiver Operating Characteristic 
Curve of \texttt{TensorRT} AI Ensemble}.  The output of our inference-optimized 
AI ensemble is used to estimate the true positive rate with a test set of 
237,663 modeled waveforms whitened with advanced LIGO data, 
and which cover a broad range of signal-to-noise ratios. The false positive rate 
is computed using a 5 year long time-shifted advanced LIGO dataset. 
The gray dashed rectangle in the left of this panel is shown in detail in the top right inset.
}
\label{fig:roc_curve}
\end{figure}

\noindent We have also computed the receiver operating characteristic (ROC) of 
our AI ensemble, shown in Figure~\ref{fig:roc_curve}. 
We computed this ROC curve using a test set of 237,663 
waveforms that cover a broad range of signal to noise ratios. 
To compute the ROC curve, we 
used an automated post-processing script that takes in the output 
of our AI ensemble, and then uses the \texttt{find\_peaks} algorithm to 
identity peaks whose width is at least 0.5 seconds long.  
As shown in Figure~\ref{fig:roc_curve}, our AI ensemble attains optimal 
true positive rate as we increase the detection threshold, or height in our  
\texttt{find\_peaks} algorithms, between 0 and 0.9998. This plot 
indicates that our AI ensemble reports, on average, one misclassification per month of searched 
data. It is worth comparing this figure to other recent studies in the literature. For instance, 
in~\cite{2021PhLB..81236029W}, it was reported that an ensemble of 2 AI models reported 
1 misclassification for every 2.7 days of searched data, and more basic AI architectures reported one misclassification for every 200 seconds 
of searched advanced LIGO data~\cite{George:2018PhLB}. For completeness, it is worth mentioning 
that the results we 
present in Figure~\ref{fig:roc_curve} differ from those we computed 
with traditional \texttt{TensorFlow} models in less than \(0.01\%\)~\cite{huerta_nat_ast}.

It remains to be seen whether adding real glitches to the training stage 
further improves the detection capabilities of our AI ensemble. We will explore 
the use of real glitches, e.g., using the catalog curated by
the Gravity Spy project~\cite{grav_spy}, to further improve 
the resilience of our AI models to noise anomalies 
through adversarial training. Having developed the required 
framework to time-shift data, in future work we will use a revised 
version of this AI ensemble to search for gravitational waves over 
entire observing run datasets. Specific future directions of work involve the 
production of software and computing methods to post-process the output data of 
our AI ensemble. At present, our AI ensemble produces about 500GB of output data
for every month of searched data. Thus, for the 5 year time-shifted advanced LIGO 
dataset we considered in this article, we post-processed 
\(5*12*500\textrm{GB}\rightarrow 30\textrm{TB}\) of 
output data by parallelizing the computing over 1216 AMD EPYC 7742 cores. 
Thus, while we can now use this method to search for gravitational waves in advanced LIGO data 
that encompass entire observing run datasets, we will introduce in future work new 
methods to quantify on the fly the sensitivity of our AI ensemble using hundreds 
of years worth of time-shifted advanced LIGO data.

\subsection{Computational Efficiency}

\begin{figure}[h!]
\includegraphics[width=\linewidth]{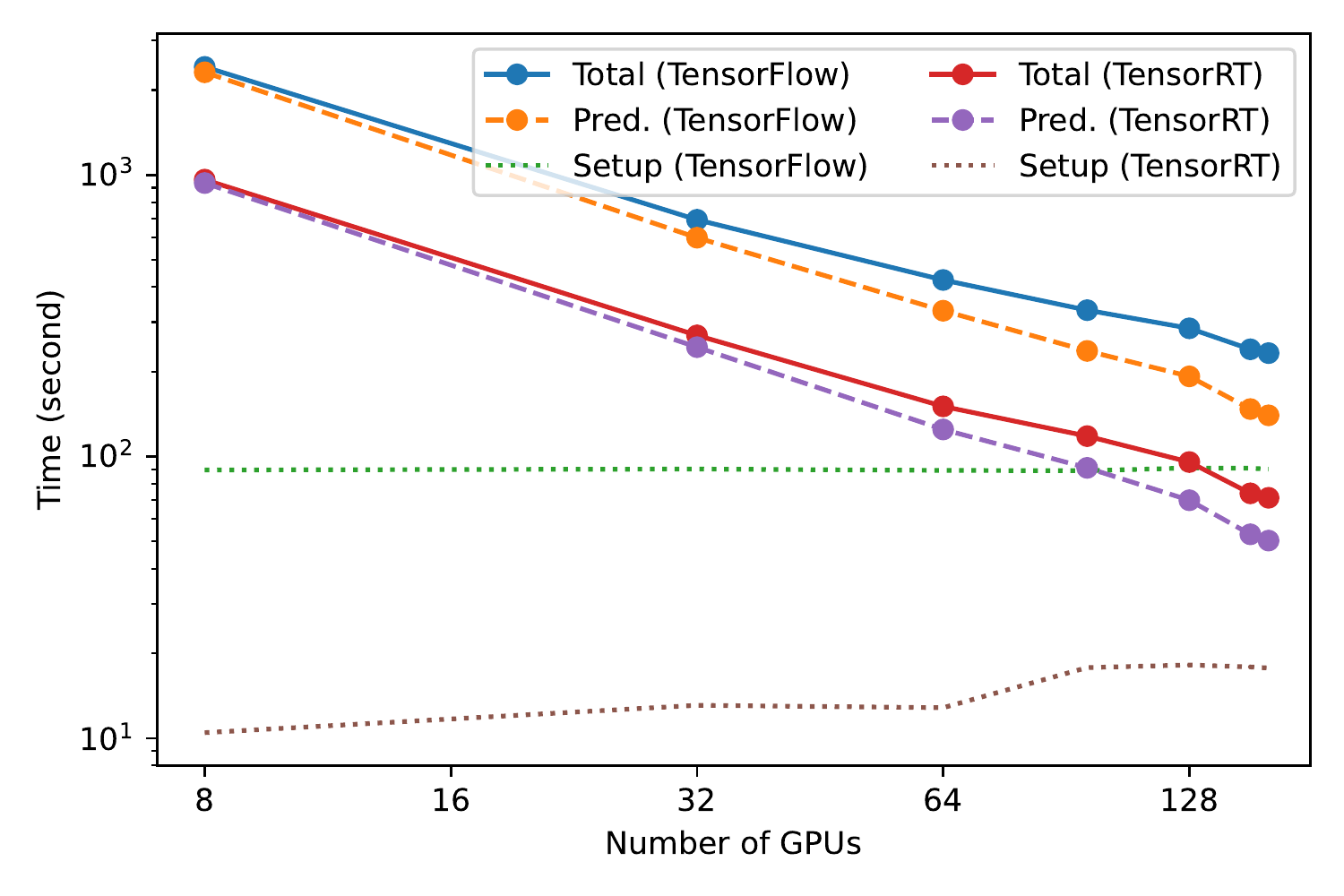}
\caption{\textbf{Scaling of Accelerated Inference in 
ThetaGPU}. \texttt{TensorRT} AI ensembles 
accelerate gravitational wave detection 
by \(3\,\textrm{fold}\) when 
compared to traditional AI ensembles (labelled as 
\texttt{TensorFlow}). \texttt{TensorRT} AI ensembles 
process an entire month of 
advanced LIGO data, including both Hanford and 
Livingstone strain data, within 50 seconds when 
AI inference is distributed over 160 NVIDIA A100 
Tensor Core GPUs in the ThetaGPU supercomputer.
}
\label{fig:thetag_scaling}
\end{figure}

\noindent We trained the 
models in our AI ensemble using distributed 
training in the Summit supercomputer. Each model 
was trained using 192 NVIDIA V100 GPUs within 2 hours. 
Thereafter, we distributed 
the inference using 160 NVIDIA A100 Tensor Core GPUs.
Figure~\ref{fig:thetag_scaling} presents 
scaling results as we distributed AI inference in 
the ThetaGPU supercomputer using both 
traditional AI models, labelled as \texttt{TensorFlow}, 
and inference-optimized AI models, labelled as 
\texttt{TensorRT}. These results show that 
our \texttt{TensorRT} AI ensemble provides a \(3X\) 
speedup over traditional AI models~\cite{huerta_nat_ast}. 
These results also 
indicate that the environment setup we used in ThetaGPU 
optimally handled I/O and data distribution 
across nodes. It is worth mentioning that these results 
were reproduced using \texttt{TensorRT} AI ensembles 
in Singularity containers, and by running our 
\texttt{TensorRT} AI ensemble natively 
on ThetaGPU using a suitable 
\texttt{Conda} environment~\cite{conda}.
Furthermore, we found that our 
\texttt{TensorRT} AI 
ensemble provides additional speedups when we 
consider larger volume datasets. We will explore the 
application of this approach for significantly larger 
datasets in the near future, and will make available these 
\texttt{TensorRT} AI models through the Data and Learning 
Hub for Science~\cite{dlhub,li147dlhub} 
so that the broader gravitational wave 
community may harness/extend/improve these AI tools 
for accelerated gravitational wave data analysis.

\begin{figure}[h!]
\includegraphics[width=\linewidth]{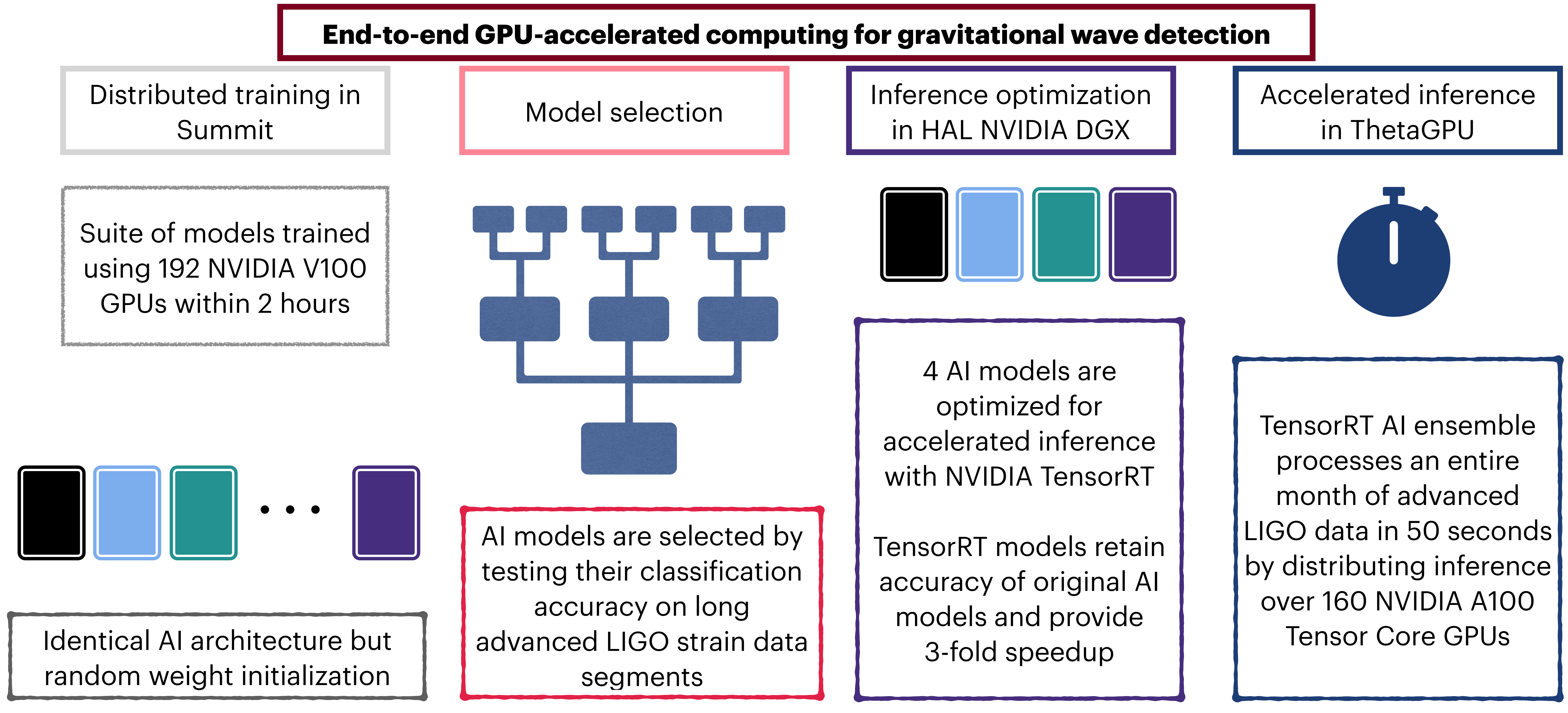}
\caption{\textbf{Convergence of AI and HPC}. 
Schematic representation of our methodology to harness 
disparate HPC platforms and data science tools to create 
optimal AI ensembles for gravitational wave detection. 
}
\label{fig:end_to_end}
\end{figure}

\noindent This study provides an exemplar 
that combines HPC systems of different scale 
to conduct accelerated AI-driven discovery, 
as shown in Figure~\ref{fig:end_to_end}. 
We showcase how to optimally use hundreds of GPUs to 
reduce time-to-insight for training (Summit) and 
inference (ThetaGPU). It is worth mentioning that 
we deliberately followed this approach, i.e., using two 
different machines for training and inference, to 
quantify the reproducibility and interoperability of 
our AI ensemble. Another important consideration is that 
we optimized our AI ensemble with \texttt{NVIDIA TensorRT} 
using an \texttt{NVIDIA DGX A100} box at the National 
Center for Supercomputing Applications. Using this 
same resource, we containerized our \texttt{TensorRT} 
AI ensemble using both Docker and Singularity. In brief, 
our methodology ensures that our AI-driven analysis is 
reproducible, interoperable and 
scalable across disparate HPC platforms.

\section{Conclusion}
\label{sec:end}

The first generation of AI models for gravitational wave 
detection exhibited great promise to accelerate 
gravitational wave discovery~\cite{geodf:2017a,George:2018PhLB}, 
and increase the science reach of 
gravitational wave astrophysics. Those models provided 
a glimpse of what may be accomplished if we were 
able to tap on the computational efficiency and 
scalability of AI. That vision is gradually coming to 
fruition by remarkable advances by multiple 
teams across the world~\cite{Nat_Rev_2019_Huerta, huerta_book,cuoco_review}. 

In this article we have described how to combine 
AI and HPC to 
accelerate the training of AI models, optimize them for 
inference, and then maximize their science throughput by 
distributing inference over tens of GPUs. 
This line of work has been explored in the context of 
AI-inference optimized applications for early warning 
systems. For instance, \texttt{PyTorch} models for 
AI forecasting of binary neutron star and black 
hole-neutron star systems were quantized to reduce 
their size by \(4X\) and accelerate their speed 
\(2.5X\) for rapid inference at the edge~\cite{Wei_quantized}. Furthermore, the combination of \texttt{TensorRT} AI 
models for data cleaning, and AI models for black hole 
detection under the umbrella of a generic inference  as a service model that leverages HPC, 
private or dedicating computing was introduced in~\cite{2021arXiv210812430G}. 
On the other hand, this work is the first in the 
literature to combine 
\texttt{TensorRT} AI models for accelerated 
signal detection with HPC at scale to process 
one month of advanced LIGO strain data from 
Hanford and Livingston within 
50 seconds using an ensemble of 4 \texttt{TensorRT} 
AI models. We have not compromised the classification 
accuracy of our models, and have found that they 
can identify all four binary black hole mergers 
previously reported 
in this data batch, namely, GW170809, GW170814, GW170818, 
and GW170823, with no misclassifications. When using a time-shifted 
advanced LIGO dataset that spans five years worth of data, we found that 
our AI ensemble reports 1 misclassification per month of searched data. This 
should be contrasted with the first generation of AI models that reported 
1 misclassification for every 200 seconds of searched data~\cite{George:2018PhLB}, and 
the other AI ensembles that reported 1 misclassifications for every 2.7 days 
of searched data~\cite{2021PhLB..81236029W}.

We are at a tipping point in gravitational 
wave astrophysics. The number of sources to be detected in the near future will 
overwhelm available and future computational resources if we 
continue to use poorly scalable and compute-intensive 
algorithms. We hope that the AI models we introduce in this 
paper are harnessed, tested, and further developed by 
the worldwide community of AI developers in gravitational wave 
astrophysics. Such an approach will provide the means 
to transform the upcoming deluge of gravitational 
wave observations into discovery at scale.

\section{Acknowledgments}
\label{ack}
\noindent P.C., E.A.H., A.K., and M.T. gratefully 
acknowledge National 
Science Foundation (NSF) awards OAC-1931561 and 
OAC-1934757. This research used resources of the Argonne 
Leadership Computing Facility, which is a DOE Office of 
Science User Facility supported under Contract DE-AC02-06CH11357.
E.A.H. gratefully acknowledges the Innovative and 
Novel Computational Impact on Theory and Experiment 
project `Multi-Messenger Astrophysics 
at Extreme Scale in Summit'. This research used 
resources of the Oak Ridge Leadership Computing Facility, 
which is a DOE Office of Science User Facility 
supported under contract no. DE-AC05-00OR22725. 
This work utilized resources supported by the 
NSF's Major Research Instrumentation program, 
the HAL cluster (grant no. OAC-1725729), 
as well as the University of Illinois at 
Urbana-Champaign. We thank \texttt{NVIDIA} for 
their continued support. 

\bibliography{book_references}
\bibliographystyle{ieeetr}

\end{document}